\documentclass[preprint,12pt]{msml2021} 

\title[Ground States of Quantum Many Body Lattice Models via Reinforcement Learning]{Ground States of Quantum Many Body Lattice Models via Reinforcement Learning}
\usepackage{times, braket, cleveref, easy-todo}
\usepackage{graphicx}
\usepackage{tabulary}
\usepackage{algorithm}
\usepackage{algorithmic}
\usepackage{comment}
\usepackage[normalem]{ulem}

\msmlauthor{\Name{Willem Gispen} \Email{wg265@cantab.ac.uk}\and
 \Name{Austen Lamacraft} \Email{al200@cam.ac.uk}\\
 \addr TCM Group, Cavendish Laboratory, University of Cambridge, J. J. Thomson Ave., Cambridge CB3 0HE, UK}

\DeclareMathOperator*{\E}{\mathbb{E}}
\newcommand*{\ba}{\vec{a}}

\newcommand*{\bs}{\vec{s}}

\newcommand*{\bbP}{{\mathbb{P}}}

\newcommand*{\cO}{{\mathcal{O}}}
\newcommand*{\cH}{{\mathcal{H}}}

\makeatletter
 \let\Ginclude@graphics\@org@Ginclude@graphics
\makeatother
\begin{document}

\maketitle

\begin{abstract}%
We introduce reinforcement learning (RL) formulations of the problem of finding the ground state of a many-body quantum mechanical model defined on a lattice. We show that stoquastic Hamiltonians -- those without a sign problem -- have a natural decomposition into stochastic dynamics and a potential representing a reward function. The mapping to RL is developed for both continuous and discrete time, based on a generalized Feynman--Kac formula in the former case and a stochastic representation of the Schr\"odinger equation in the latter. We discuss the application of this mapping to the neural representation of quantum states, spelling out the advantages over approaches based on direct representation of the wavefunction of the system.
\end{abstract}



\section{Introduction}

Finding the ground state of a quantum mechanical system involves finding the lowest eigenvalue of the Hamiltonian operator that describes the system. For many body systems, this is a problem that grows exponentially harder as the number of particles increases, and finding tractable approaches has been the principal computational challenge of quantum mechanics since its inception. 

When studying a non-relativistic system of electrons and nuclei, a natural starting point is a realistic Hamiltonian in continuous space with Coulomb interactions. Often, however, \emph{model Hamiltonians} are used instead: for example, the singularity of the Coulomb potential describing the electron-ion attraction is often replaced with a less singular \emph{pseudopotential} in quantum chemistry calculations. Another type of simplification, more common in strongly correlated physics, is to study \emph{lattice models} where particle locations are discretized. Although less realistic, this simplification often allows the treatment of larger systems where qualitative different thermodynamic phases become apparent.

In recent years, progress in deep learning has opened up new fronts to attack the many body problem by parameterizing the wavefunction describing the ground state using neural networks (e.g.\ \cite{carleo_solving_2017,choo_two-dimensional_2019, pfau_ab-initio_2019}). In this work we introduce a new neural approach to lattice models using an alternative formulation based on Reinforcement Learning (RL), in which the optimal policy matches a stochastic process describing the ground state probability distribution. \cite{barr_quantum_2020} applied the same perspective to continuum models using the Feynman--Kac formula and tools of stochastic calculus. We will present the analogous theory for the lattice case, and introduce algorithms to learn the optimal policy in this setting.

Our formulation draws together concepts from many-body physics, control theory, and reinforcement learning. The next Section introduces the necessary background, before \Cref{sec:rl4qm} presents the RL formulation of quantum states. \Cref{sec:neural} describes the application of this formulation to the neural representation of quantum states, before we present our conclusions in \Cref{sec:conclusion}.

\section{Theoretical Background}

In this Section we will introduce the models to be studied and two stochastic representations: in continuous time using a generalized Feynman--Kac formula, and in discrete time using the Schr\"odinger equation directly. We then describe the class of linearly solvable Markov decision problems introduced in \cite{todorov2006linearly}, which form the basis of our mapping to RL. The connections between RL and quantum physics are well established in the case of continuous state spaces: see \cite{barr_quantum_2020}
 and references therein.

\subsection{Models}

For illustration purposes, we restrict ourselves to the simplest family of lattice models describing a system of $N$ spins taking values $Z_i=\pm 1$. \footnote{The term \emph{lattice gas} is sometimes used, where the interpretation is that sites in the lattice are either occupied or empty.} The Hamiltonians act on the $2^N$-dimensional space $\mathbb{C}^2\otimes \mathbb{C}^2 \otimes \cdots \mathbb{C}^2$ and are expressed in terms of Pauli spin matrices $\{X_i, Y_i, Z_i\}$ 
\begin{equation}
    X_i = \begin{pmatrix}
    0 & 1 \\
    1 & 0
    \end{pmatrix}_i\qquad 
    Y_i = \begin{pmatrix}
    0 & -i \\
    i & 0
    \end{pmatrix}_i\qquad
    Z_i = \begin{pmatrix}
    1 & 0 \\
    0 & -1
    \end{pmatrix}_i
\end{equation}
acting on spin $i$. We consider two examples, the \textbf{transverse field Ising model}
\begin{equation} \label{eq:ising}
      H_\text{Ising} = -J \sum_{\braket{i,j}} Z_{i} Z_{j}
        - h \sum_{i} X_i, 
\end{equation}
and the \textbf{XXZ model} 
\begin{equation} \label{eq:xxz}
      H_{XXZ} = -\sum_{\braket{i,j}} \left[J Z_{i} Z_{j} + J_\perp\left(X_iX_j + Y_iY_j\right)\right],
\end{equation}    
where $\braket{i,j}$ denotes a sum over nearest neighbors on the lattice. The case $J=J_\perp$ is known as the XXX or Heisenberg model \citep{auerbach2012interacting}.
    
\subsection{Connection to stochastic dynamics}

\begin{figure}
    \centering
    \includegraphics[height=0.43\textwidth]{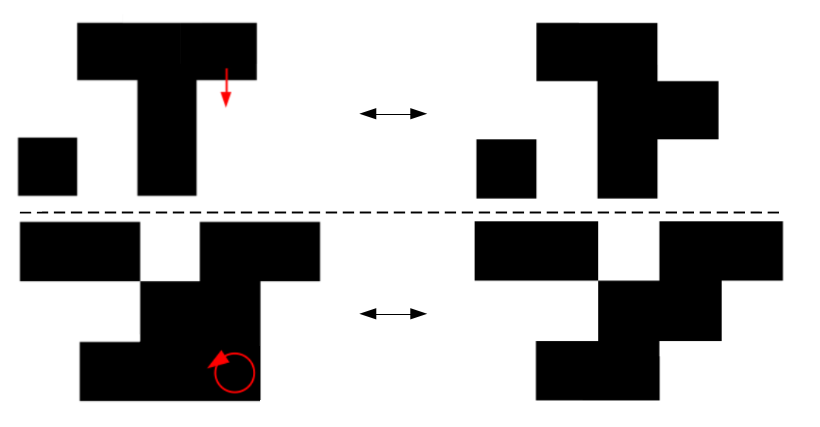}
    \caption{Transitions in the simple exclusion process (above) and the Ising model (below). In the simple exclusion process, a transition is a move of a particle to a non-occupied (white) lattice site. For the Ising model, a transition is a flip of spin up (black) to spin down (white) or the other way around.}
    \label{fig:isingandxy}
\end{figure}

Particular limits of the above models can be described in terms of stochastic dynamics. We illustrate this connection in the case of the XXX model by observing that in the space of two neighboring spins
\begin{equation}
    \left[Z_{i} Z_{j} + X_iX_j + Y_iY_j\right]-1= \begin{pmatrix}
    0 & 0 & 0 & 0 \\
    0 & -1 & 1 & 0 \\
    0 & 1 & -1 & 0 \\
    0 & 0 & 0 & 0
    \end{pmatrix}_{{ij} }.
\end{equation}
This may be interpreted as a transition rate matrix for a Markov process with transitions $\circ \bullet\longleftrightarrow \bullet \circ$, where filled and empty circles represent spin up and spin down states. Apart from an additive constant, the operator $-H_\text{XXX}$ for $J>0$ is then a transition rate matrix of the \textbf{symmetric simple exclusion process}, describing a lattice gas of diffusing particles (\Cref{fig:isingandxy}). 
\begin{equation}
-H_\text{XXX} = {\overbrace{J\sum_{\braket{i,j}}\left[Z_{i} Z_{j} + X_iX_j + Y_iY_j-1\right]}^{\equiv\Gamma_{\text{SEP}}}} + JN.
\end{equation}
A valid transition matrix has columns summing to zero; by symmetry of the Hamiltonian this is also true for the rows. The ground state of $H_\text{XXX}$ corresponds to the dominant (maximal) eigenvector of the transition matrix with eigenvalue 0. This eigenvector describes a uniform distribution of particle configurations.

When $J\neq J_\perp$ the Hamiltonian no longer has a simple stochastic interpretation, having the form
\begin{equation}\label{eq:stoq}
 H_\text{XXZ} = -\Gamma_{\text{SEP}} + V,
\end{equation}
where $V$ is diagonal but not constant. Hamiltonians of this form are called \emph{stoquastic}, and we will refer to the Markov process defined by $\Gamma$ as the \emph{passive dynamics}, for reasons that will become clear. In terms of the matrix elements of the Hamiltonian, the potential is $V_{ss'}=V(s)\delta_{ss'}$, where
\begin{equation} \label{eq.potential.from.H}
  V(\bs) = H_{\bs\bs} + \sum_{\bs' \neq \bs} H_{\bs \bs'}.  
\end{equation}
Moreover, note that \eqref{eq:stoq} implies that $H_{\bs\bs'}\leq0$ for $\bs\neq\bs'$.

As a second example, consider $H_\text{Ising}$, which is stoquastic for $h>0$ with very simple transition rates
\begin{equation}
 \Gamma_{{\bs}\to{\bs'}} = h,
\end{equation}
if the two spin configurations $s$ and $s'$ differ at only one site (\Cref{fig:isingandxy}). From \Cref{eq.potential.from.H}, the potential is
\begin{align*}
V(\bs)
    &= -\sum_{{\bs'}\neq{\bs}} \Gamma_{{\bs}\to{\bs'}} + H_{ss}\\
    &= -hN -  J \sum_{\braket{i,j}} {s_i} {s_j}.
\end{align*}
%

A key property of stoquastic Hamiltonians is the following: if the passive dynamics is ergodic, then the ground state wavefunction is positive \citep{bravyi_monte_2015}. This is a consequence of the Perron-Frobenius theorem, and explains why stoquastic Hamiltonians are `free of the sign problem': the ground state wavefunction does not change sign.

\subsection{Feynman--Kac formula}\label{sec:fk}

For a stoquastic Hamiltonian, a representation of the ground state in terms of the passive dynamics exists, in the form of the Feynman--Kac formula. Although this is more familiar in the context of continuous state spaces where the underlying stochastic dynamics is Brownian motion, we will require a discrete analog.

First, note that the ground state $\varphi_0$ of the Hamiltonian is the asymptotic solution of the imaginary time Schr\"odinger equation 
\begin{equation} \label{eq.imaginary.schrodinger}
    \partial_t \varphi(t) = - H \varphi(t)
\end{equation}
$\varphi(t)\to e^{-E_{0}  t}\varphi_0$ as $t\to\infty$, where $E_0$ is the ground state energy \citep{goldberg_integration_1967}.

The Feynman--Kac formula gives the solution of the imaginary time Schr\"odinger equation with a \emph{terminal} condition at some time $t+T$.
For a stoquastic Hamiltonian with passive rates $\Gamma$ and potential $V$, the Feynman--Kac formula states that the solution at the earlier time $t$ can be expressed as an expectation \citep{rogers_diffusions_2000}
\begin{equation}
 \varphi({\bs_t}, {t}) = \E_\bbP
 \left[\exp\left(-\int_{t}^{t+T} V({\bs_{t'}}) dt'\right)\varphi({\bs_{t+T}}, {t+T})\right].
\end{equation}
Here $\varphi({\bs_{t+T}}, {t+T})$ is the terminal condition, and the expectation is over trajectories following the passive dynamics $\Gamma$ starting at $s_t$ at time $t$.
Since $\varphi({\bs_t}, {t}) = e^{-E_0 t}\varphi_0({\bs_t})$ is an exact solution of the imaginary time Schr\"odinger equation, the Feynman--Kac formula implies
\begin{equation} \label{eq.fk.ground}
\varphi_0(\bs_t) = \E_\bbP
 \left[\exp\left(-\int_{t}^{t+T} \left (V({\bs_{t'}}) {+} E_0 \right ) dt' \right)\varphi_0({\bs_{t+T}})\right].
\end{equation}
In \eqref{eq.fk.ground} the role of the potential is to modify the measure over trajectories relative to that of the passive dynamics. It is perhaps not obvious, but nonetheless true, that the modified measure describes the trajectories of a different Markov process. In \Cref{sec:transform.fk}, we formulate the problem of learning the Markov process that correctly describes the trajectories in the Feynman--Kac formula as a Reinforcement Learning problem.

\subsection{Stochastic representations of the Schr\"odinger equation} \label{sec:stoch}

The Feynman--Kac formula gives a natural stochastic representation of the ground state in terms of a \emph{continuous} time Markov chain. As discussed in \cite{neirotti1996monte}, the ground state of a stoquastic Hamiltonian can also be characterized by the stationary state of a \emph{discrete} time Markov chain. Inspired by this result, we give two stochastic representations of the time-independent Schr\"odinger equation
\begin{equation} \label{eq:schrodinger}
  \sum_{\bs'} H_{\bs \bs'} \varphi_0(\bs') = E_0 \varphi_0(\bs).  
\end{equation}
%
For stoquastic Hamiltonians, the ground state wavefunction $\varphi_0$ is the unique positive eigenfunction.

For the first representation, fix $C > \max_{\bs} H_{\bs\bs}$. For any such constant, we can rewrite the Schr\"odinger equation \eqref{eq:schrodinger} as
\begin{equation}
  \varphi_0(\bs) = \frac{1}{C-E_0} \left [ (C - H_{\bs\bs}) \varphi_0(\bs) - \sum_{\bs'\neq\bs} H_{\bs \bs'} \varphi_0(\bs') \right].  
\end{equation}
Since $C > \max H_{\bs\bs}$ and $H_{\bs\bs'}<0$ for $\bs'\neq\bs$, we can turn this into a stochastic representation
\begin{equation} \label{eq:stochastic.schrodinger.1}
 \varphi_0(\bs) = \frac{Z_1(\bs)}{C-E_0} \E_{\bs'\sim p_1(\cdot|\bs)} \left [ \varphi_0(\bs') \right].
\end{equation}
Here the transition probabilities $p_1(\bs'|\bs)$ are
\begin{equation}
    p_1(\bs'|\bs) =
    \begin{cases} 
        \left ( C - H_{\bs\bs} \right) / Z_1(\bs) & \bs' = \bs \\
        - H_{\bs\bs'} / Z_1(\bs) & \bs' \neq \bs
    \end{cases}
\end{equation}
with $Z_1(\bs) = C - H_{\bs\bs} - \sum_{\bs'\neq\bs} H_{\bs \bs'}$. 

For the second representation, we rewrite the Schr\"odinger equation slightly differently as
\begin{equation}
   \varphi_0(\bs) = -\frac{1}{H_{\bs\bs} - E_0} \sum_{\bs'\neq\bs} H_{\bs\bs'} \varphi_0(\bs') 
\end{equation}
This leads to the stochastic representation
\begin{align}\label{eq:stochastic.schrodinger.2}
 \varphi_0(\bs) &= \frac{Z_2(\bs)}{H_{\bs\bs} - E_0} \E_{\bs'\sim p_2(\cdot|\bs)} \left [ \varphi_0(\bs') \right]\\
 p_2(\bs'|\bs) &=
            \begin{cases} 
        0 &  \bs' = \bs \\
        - H_{\bs\bs'} / Z_2(\bs) & \bs' \neq \bs
    \end{cases}
\end{align}
with $Z_2(\bs) = -\sum_{\bs'\neq\bs} H_{\bs\bs'}$. Note that in this second formulation the system always changes state at each step of the process. In both cases, the formulation requires knowledge of the ground state energy $E_0$.


\subsection{Linearly solvable reinforcement learning problems}

The stochastic formulation of the Schr\"odinger equation given in the previous section may be reinterpreted as a linearly solvable Markov decision problem of the type introduced in \cite{todorov2006linearly}. 

Todorov considers the maximum entropy reinforcement learning paradigm, where stochastic policies are encouraged by complementing the reward function $r(\bs, \ba)$ with the entropy of the policy relative to some reference policy $p$.\footnote{For details see Appendix \ref{sec:maxent.rl}.} As is usual in RL, a problem setting can be summarized and studied by its Bellman equation. Two examples of (soft) Bellman equations are
\begin{subequations}
\begin{equation} \label{eq.u.soft.bellman}
 U^*(\bs) = R^* + \log \E_{\ba \sim p(\cdot|\bs)} \left [\exp \left ( r(\bs,\ba) + U^*(\ba(\bs)) \right )\right ]
\end{equation}
\begin{equation} \label{eq.u.soft.bellman.terminal}
 U^*(\bs) = \log \E_{\ba \sim p(\cdot|\bs)} \left [\exp \left ( r(\bs,\ba) + U^*(\ba(\bs)) \right )\right ].
\end{equation}
\end{subequations}
Here $r(\bs, \ba)$ is the reward following the action $\ba$ in the state $\bs$ and $R^*$ is a constant describing the average reward of the optimal policy. $U^*$ is the optimal state-value function and $p$ is the reference policy. Information implicitly contained in these soft Bellman equations is the objective function and time horizon. While \eqref{eq.u.soft.bellman} describes an infinite time horizon, \eqref{eq.u.soft.bellman.terminal} describes a problem with terminal states. Both consider an undiscounted setting where the objective function is simply the sum (or average) of received rewards.

\cite{todorov_efficient_2009} showed that for action-independent rewards, both soft Bellman optimality equations \eqref{eq.u.soft.bellman} and \eqref{eq.u.soft.bellman.terminal} can be transformed into linear equations. Consider action-independent rewards $r(\bs,\ba) = r(\bs)$. The key is an exponential transformation to the `desirability' $z$ of a state
\begin{equation}
 z(\bs) = \exp(U^*(\bs)).
\end{equation}
By simply exponentiating the soft Bellman optimality equations \eqref{eq.u.soft.bellman} and \eqref{eq.u.soft.bellman.terminal}, we obtain
\begin{subequations}
\begin{equation} \label{eq.todorov.desirability}
 z(\bs) = e^{r(\bs) + R^*} \E_{\ba\sim p(\cdot|\bs)} \left [ z(\ba(\bs)) \right],    
\end{equation}
\begin{equation} \label{eq.todorov.desirability.2}
 z(\bs) = e^{r(\bs)} \E_{\ba\sim p(\cdot|\bs)} \left [ z(\ba(\bs)) \right],    
\end{equation}
\end{subequations}
which are indeed linear equations for the desirability $z$. { Note that the introduction of discounting results in nonlinear equations, breaking the connection with the Schr\"odinger equation.} 




\section{Reinforcement learning for quantum ground states} \label{sec:rl4qm}

Our first contribution is to show how to transform the stochastic representations of \Cref{sec:fk,sec:stoch} to reinforcement learning problems. The key step is a logarithmic transformation that transforms the stochastic representation into a Bellman equation, and represents the inverse of Todorov's approach.

\subsection{Transforming the Feynman-Kac representation}
\label{sec:transform.fk}

The Feynman--Kac representation of the ground state \Cref{eq.fk.ground} may be formulated as a continuous time RL problem. In this section, we give an idea of the proof, see \Cref{sec:fkbellman} for details.

\begin{figure}
    \centering
    \includegraphics[width=0.6\textwidth]{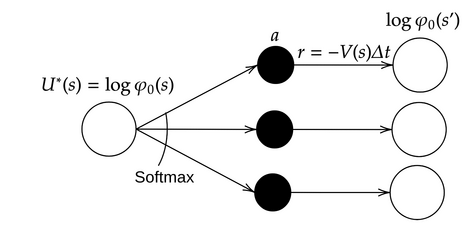}
    \caption{The Feynman--Kac formula seen as a Bellman backup diagram.}
    \label{fig:bellman}
\end{figure}

Consider a small time step $\Delta t$ for the Feynman--Kac characterization of the ground state \eqref{eq.fk.ground}
\begin{equation}
\varphi_0(\bs_{t}) 
    = e^{-V(\bs)\Delta t+E_0 \Delta t} \E_{\bbP} \left [  \varphi_0 (\bs_{t+\Delta t})) \right] +\cO(\Delta t^2).
\end{equation}
Comparison with \eqref{eq.todorov.desirability} gives a reward $r(\bs) = - V(\bs) \Delta t$. Furthermore, the ground state wavefunction takes the role of the desirability $z$, so that we can make the identification
\begin{equation} \label{eq:wave.from.u}
    \log \varphi_0(\bs) = U^*(\bs).
\end{equation}
This identification is visualized in \Cref{fig:bellman} as a Bellman backup diagram. Taking the continuous time limit, this leads to the following maximum entropy RL problem. The control is provided by transition rates $\Gamma^{\theta}$ that drive a continuous time Markov chain for the state $\bs_t$ of the system. We seek the optimal rates $\Gamma^*$
\begin{align}
\Gamma^* &= \mathop{\text{argmax}}_{\Gamma^\theta} R\\
\label{eq:holland.reward.discrete}
  R &= -\lim_{T\to\infty} \frac{1}{T} \E_{s_{0:T}} \left[\int_0^T \left(V(\bs_t) +    \mathcal{H} (\bs_t) \right) dt \right], 
\end{align}
where the entropy term $\mathcal{H}$ is given by
\begin{equation} \label{eq:entropy}
    \mathcal{H} (\bs_t) = \sum_{\bs'\neq \bs_t} \left(
    {\Gamma_{\bs_t\to \bs'}} 
    - \Gamma^{\theta}_{\bs_t\to \bs'}
    + \Gamma^{\theta}_{\bs_t\to \bs'} 
        \log\left(\frac{\Gamma^{\theta}_{\bs_t\to \bs'}}{\Gamma_{\bs_t\to \bs'}}\right) \right),
\end{equation}
and $\Gamma_{s\to s'}$ denotes the passive dynamics of the system
\begin{equation} \label{eq.passive.dynamics}
 \Gamma_{\bs\to \bs'} =
\begin{cases}
    - H_{\bs \bs'} & \bs\neq \bs \\
    \sum_{\bs' \neq \bs} H_{\bs \bs'} & \bs=\bs' \\
\end{cases}
\end{equation}
The expectation in the objective \eqref{eq:holland.reward.discrete} is over trajectories $s_{0:T}$ following the parameterized dynamics $\Gamma^\theta$.

The optimal policy strikes a balance between steering towards states $\bs$ with low potential energy $V(\bs)$ and staying close to the passive dynamics $\Gamma$ of \Cref{eq.passive.dynamics}. In physical terms, the objective \eqref{eq:holland.reward.discrete} minimizes the sum of potential energy $V(\bs)$ and kinetic energy $\cH$. { The optimal rates $\Gamma^*$ reproduce the Feynman--Kac measure on the space of trajectories . Note that \eqref{eq:holland.reward.discrete} involves an expectation over the parameterized Markov process. In the case of continuous state spaces, optimizing this expectation is straightforward using the reparameterization trick. The discrete case requires a different approach as described below.} 

\subsection{Transforming the Schrodinger representations}

In a similar way, the stochastic representations of the Schr\"odinger equation from \Cref{sec:stoch} can be mapped to RL in discrete time. The linear equations for the desirability \Cref{eq.todorov.desirability,eq.todorov.desirability.2} will be identified with the stochastic representations \Cref{eq:stochastic.schrodinger.1,eq:stochastic.schrodinger.2} of the Schr\"odinger equation.

The first representation \Cref{eq:stochastic.schrodinger.1} maps to the soft Bellman equation \Cref{eq.u.soft.bellman} i.e.\
\begin{equation}
     U^*(\bs) = R^* + \log \E_{\ba \sim p_1(\cdot|\bs)} \left [\exp \left ( r(\bs) + U^*(\ba(\bs)) \right )\right ].
\end{equation}
The identification is $\log \varphi_0(\bs) = U^*(\bs)$ as before, but now with reward function
\begin{equation}
   r(\bs) = \log Z_1(\bs)
\end{equation}
and $R^* = -\log(C - E_0)$.
Therefore, the ground state problem is also equivalent to a discrete time RL problem with an infinite time horizon.

The second representation \eqref{eq:stochastic.schrodinger.2} maps to the soft Bellman equation \eqref{eq.u.soft.bellman.terminal}
\begin{equation}
 U^*(\bs) =  \log \E_{\ba\sim p_2(\cdot|\bs)} \left [\exp \left ( r(\bs) + U^*(\ba(\bs)) \right )\right ]
\end{equation}
with reward
\begin{equation}
    r(\bs) = \log \left ( \frac{Z_2(\bs)}{H_{\bs\bs}-E_0} \right).
\end{equation}
This Bellman equation describes a RL problem with terminal states. In this type of RL problem, there is no fixed time horizon, but exploration stops when one of terminal states is reached. The reward is the total received until the terminal state is reached, which means terminal states necessarily have state-value $U^*=0$.

We are free to choose the terminal states. Choosing a terminal state, i.e.\ setting $U^*(\bs)$ to zero, is equivalent to forcing $\varphi_0(\bs)=1$ because of the logarithmic transformation $U^*= \log\varphi_0$. Therefore, a choice of terminal state entails a choice of normalization for the wavefunction. A natural candidate for the terminal state is the classical ground state, i.e.\ the state $\bs$ for which $H_{\bs\bs}$ is minimal. Another choice that could be advantageous computationally is a state $\bs$ with minimum symmetry. This is because states $\bs$ and $\bs'$ that are related by a symmetry (such as translation) must obey $\varphi_0(\bs) = \varphi_0(\bs')$, so that choosing a minimum symmetry state gives you many other terminal states for free.



\section{Application: neural quantum states} \label{sec:neural}

In this section, we describe how our mapping of the ground state problem to RL leads to new optimization techniques for neural quantum states. Usually, the quantum wavefunction is parameterized by a neural network $\varphi(\bs) =\mathsf{NN}(\bs)$. In contrast, almost all deep reinforcement learning algorithms obtain a neural approximation of either the state-value function $U^*(\bs)$ or the action-value function $Q^*(\bs, \ba)$. Given a neural approximation of the state-value function $U^*(\bs)$, we immediately obtain a neural approximation of the ground state wavefunction
\begin{equation}
 \varphi_0(\bs) = \exp(U^*(\bs))    
\end{equation}
because this is the transformation \eqref{eq:wave.from.u} we used to derive the reinforcement learning formulations.

Given a neural approximation of the action-value function $Q^*(\bs, \ba)$, we can also obtain a neural quantum state. For this we use the following relation between the state-value function $U^*$ and the action-value function $Q^*$
\begin{equation}
 U^*(\bs) = \log \E_{\ba \sim p(\cdot|\bs)} \left [\exp \left (  Q^*(\bs, \ba) \right) \right ].
\end{equation}
Therefore, a neural approximation of $Q^*$ gives a neural approximation of the ground state wavefunction
\begin{equation} \label{eq:wave.from.q}
 \varphi_0(\bs) = \E_{\ba \sim p(\cdot|\bs)} \left [\exp \left (  Q^*(\bs, \ba) \right) \right ].
\end{equation}

{

}

\subsection{Advantages of reinforcement learning neural quantum states}
Most papers on Neural Quantum States use either Variational Monte Caro (e.g.\ \cite{han_solving_2019, hermann_deep_2020, kessler_artificial_2019, saito_solving_2017}) or Stochastic Reconfiguration (e.g.\ \cite{choo_two-dimensional_2019, nomura_restricted_2017, pfau_ab-initio_2019}) to optimize neural quantum states. Our optimization method using reinforcement learning could be preferable for the following reasons.

Firstly, our approach needs less data for one update step. For example, Variational Monte Carlo needs to calculate the variational energy for each update step. This means that, given a parametrized wavefunction $\varphi$, samples $\bs\propto \varphi^2$ need to be produced and the so-called local energy
\begin{equation}
    \frac{H\varphi}{\varphi}(\bs) = \frac{1}{\varphi(\bs)} \sum_{\bs'\neq\bs} H_{\bs\bs'} \varphi(\bs')
\end{equation}
needs to be averaged over these samples. Therefore, for each update step, new samples need to be generated, and for each sample, the neural wavefunction $\varphi$ needs to be evaluated $\cO(N)$ times, where $N$ is the number of spins in the system. In contrast, in a Q-learning approach, the action-value function $Q$ can be optimized using off-policy pairs of states $\bs$ and actions $\ba$. The fact that only pairs of states and actions are needed means that the neural action-value function $Q$ only needs to evaluated $\cO(1)$ times per sample $\bs$. The fact that off-policy data can be used means that generated samples can be reused for different update steps. So one update step can be completed more efficiently using our reinforcement learning approach.

Secondly, our approach can more efficiently produce samples from the neural ground state. To see this, note that such samples are usually generated using the Metropolis--Hastings algorithm. In the Metropolis--Hastings algorithm, the proposal distribution of moves greatly influences its speed of convergence. All reinforcement learning algorithms produce a policy that prefers states of high state-value. In our formulation, that translates to a policy that points towards states of high ground state probability. { In the next section, we show how this leads to a more efficient proposal distribution. 
}


{
\subsection{Experiments}
To demonstrate the application to neural quantum states, we consider the two-dimensional transverse Ising model (\Cref{eq:ising}). Training details can be found in \Cref{sec:hparams} and the code at \ifanonsubmission \href{https://github.com/[anonymized]}{https://github.com/[anonymized]}\else \href{https://github.com/WillemGispen/Lattice-QuaRL}{https://github.com/WillemGispen/Lattice-QuaRL} \fi.

We use Soft Q-learning \citep{haarnoja_soft_2018} with discrete space adaptations \citep{christodoulou_soft_2019} to approximate the optimal action-value function $Q^*$ with a convolutional neural network. As described before, this leads via \eqref{eq:wave.from.q} to a neural representation of the ground state wavefunction. One advantage of Christodoulou's discrete space adaptations is that all action-values $Q^*(\bs, \ba)$ for a specific state $\bs$ are obtained with one neural network call, so that the neural quantum state \eqref{eq:wave.from.q} may be calculated simply by weighting the sum of this output.

For square lattices up to $6\times 6$, we can compare the variational energy of our optimized wavefunctions with exact diagonalization \citep{hamer_finite-size_2000}. The numerically exact ground state energy of the $6\times 6$ square Ising model, at $h=1$ and $J=0.32758$, is $E_0 = -1.06375$ \citep{hamer_finite-size_2000}. All our methods come within $0.1\%$ of this value, obtaining $-1.0628 (0.09\%)$, $-1.0632 (0.05\%)$, and $-1.0633 (0.04\%)$ for the continuous, non-terminal, and terminal states formulations, respectively. These errors should not be viewed as the limits of each method, but purely as a demonstration that each method can obtain the ground state with reasonable accuracy.

To produce samples from the ground state, we use Metropolis--Hastings with a variety of proposal distributions. The standard proposal distribution is uniform: a single random spin flip is proposed per Metropolis--Hastings step. Using our learned action-values, we can instead propose single spin flips according to $\exp(Q(\bs, \ba))$. Finally, we can sample multiple spin flips from $\exp(Q(\bs, \ba))$ and propose to flip those all at once.

To quantify how efficiently the different proposal distributions sample our optimized $6\times6$ Ising wavefunctions, we measure the autocorrelation of the potential energy. This autocorrelation decays approximately exponentially $\propto \exp(-t/\tau)$. Sampling single spin flips from $\exp(Q(\bs, \ba))$ instead of uniformly approximately halves $\tau$. Sampling $6$ spin flips from $\exp(Q(\bs, \ba))$ results in a $6\times$ speedup of $\tau$. Sampling $10$ spin flips from $\exp(Q(\bs, \ba))$ for a $10\times10$ Ising model (trained with the terminal states method) results in a $10\times$ speedup of $\tau$. These `speedups' refer to comparisons with a single uniformly generated spin flip and are comparable for all our methods. Although our measurements of $\tau$ are only rough indications for the true autocorrelation or mixing times, they suggest the following: sampling $\sqrt{N}$ spin flips from $\exp(Q(\bs, \ba))$ for an Ising system with $N$ spins may reduce autocorrelation/mixing times from $\cO(N)$ to $\cO(\sqrt{N})$. Anyhow, it seems clear that using $\exp(Q(\bs, \ba))$ to propose spin flips can signicantly reduce autocorrelation/mixing times. This could be particularly beneficial when autocorrelation/mixing times are long, such as for large systems or when random actions have very low acceptance rates.

}

\section{Conclusion} \label{sec:conclusion}

In this paper we have introduced three different reinforcement learning formulations of the problem of finding the ground state of a quantum lattice model, as summarized in \Cref{tab:overview}.

Even before doing {extensive} experiments, we can note some relative advantages and disadvantages of the different RL formulations. Firstly, discrete time RL settings are more commonly studied than continuous time ones, so it is easier to find efficient RL algorithms to solve the discrete time formulations. A disadvantage of the third formulation is that the reward contains the ground state energy $E_0$. Still, an estimate for the ground state energy may be obtained from the variational energy, which must be calculated anyway for validation purposes. Moreover, sparse terminal states are challenging, although they can be dealt with \citep{agostinelli2019solving}. 

\begin{table}[]
\begin{tabular}{l|lll}
Basis                  & Time       & Time horizon    & Reward                      \\
\hline
Feynman-Kac formula    & Continuous & Infinite        & $V(s)$                      \\
Schr\"odinger equation & Discrete   & Infinite        & $\log Z_1 (\bs) $           \\
Schr\"odinger equation & Discrete   & Terminal states & $\log \left ( \frac{Z_2(\bs)}{H_{\bs\bs} - E_0} \right)$
\end{tabular}
\caption{Overview of the three different reinforcement learning formulations of the ground state problem.}
\label{tab:overview}
\end{table}

Future experiments will explore the performance of our three approaches relative to each other and existing neural representations of quantum states. As well as being the basis of methods in their own right, the learned process described by the rates $\Gamma_\theta$ could be used for importance sampling of the Feynman--Kac formula, which will be optimal (i.e. with zero variance) for $\Gamma_\theta=\Gamma^*$.


\acks{We thank Ariel Barr for early discussions of this work and collaboration on a related project. AL acknowledges support from EPSRC Grant No. EP/P034616/1.}

\bibliography{msml2021}

\appendix


\section{Maximum entropy reinforcement learning}
\label{sec:maxent.rl}
This section is based on \cite{galashov_information_2019, haarnoja_reinforcement_2017, sutton_reinforcement_2018, todorov_efficient_2009}.

The usual reinforcement learning objective of maximizing average reward leads to deterministic policies. This can lead to instabilities in reinforcement learning algorithms. One attempt to alleviate this is to extend the reward with a term that promotes policies that are as random as possible. This strategy leads to the theory of maximum entropy reinforcement learning. For reinforcement learning problems, the advantages range from improved stability, exploration, and transfer learning \citep{Levine:2018aa}.

Consider an reinforcement learning agent using a stochastic policy $\pi$. This means that the agent, observing state $\bs$, chooses action $\ba$ with probability $\pi(\ba|\bs)$. The entropy regularization is ensured by complementing the reward $r(\bs, \ba)$ with an entropy term
\begin{equation}
 r(\bs, \ba) \to r(\bs, \ba) + \alpha \cH(\pi(\cdot|\bs) || p(\cdot|\bs)).
\end{equation}
Here $\alpha$ is called the `temperature': it controls the importance of the entropy term. $\cH(\pi(\cdot|\bs) || p(\cdot|\bs))$ is the relative entropy\footnote{The negative relative entropy is also called the Kullback-Leibler (KL) divergence, and this regularization strategy is therefore sometimes called KL-regularization. If the reference policy $p$ is uniform, then the relative entropy reduces to the Shannon entropy.} of the policy $\pi$ with respect to some reference policy $p$
\begin{equation}
 \cH(\pi(\cdot|\bs) || p(\cdot|\bs)) = -\E_{\ba\sim p(\cdot|\bs)} \left[ \frac{d\pi}{dp} \log \left ( \frac{d\pi}{dp} \right ) (\ba|\bs) \right ]
\end{equation}
and $\frac{d\pi(\cdot|\bs)}{dp(\cdot|\bs)}$ is the Radon--Nikodym derivative of $\pi$ with respect to $p$. Although the temperature $\alpha$ is an important parameter in maximum entropy reinforcement learning, { $\alpha=1$ is fixed by our requirement that the Bellman equation maps to a linear Schr\"odinger equation (\cite{todorov_efficient_2009})}. Thus, for an infinite time horizon, the entropy-regularized objective is to maximize
\begin{equation} \label{eq:merl.objective}
   R^{\pi} = \lim_{T\to\infty} \frac{1}{T} \E_{\bs_{0:T}} \left[ \sum_{t=0}^T r(\bs_t, \ba_t) + \cH(\pi(\cdot|\bs_t) || p(\cdot|\bs_t)) \right]
\end{equation}
where the expectation is over trajectories $\bs_{0:T}$ of the policy $\pi$.

The value function $U^*$ and the action-value function $Q^*$ are changed accordingly. Therefore, the usual Bellman optimality equations are replaced by their `soft' counterparts: the soft Bellman optimality equations (again for the example of an infinite time horizon)
\begin{subequations}
\begin{equation} \label{eq.u.soft.bellman.app}
 U^*(\bs) = R^* + \log \E_{\ba \sim p(\cdot|\bs)} \left [\exp \left ( r(\bs,\ba) + U^*(\ba(\bs)) \right )\right ]
\end{equation}
\begin{equation} \label{eq.q.soft.bellman.app}
 Q^*(\bs, \ba) = R^* + r(\bs, \ba) + \log \E_{\ba' \sim p(\cdot|\ba(\bs))} \left [\exp \left (  Q^*(\ba(\bs), \ba') \right) \right ]
\end{equation}
\end{subequations}
where the expectations are according to the reference policy $p$. The constant $R^*$ describes the average reward of the optimal policy. { When there are terminal states, these soft Bellman optimality equations become
\begin{subequations}
\begin{equation} \label{eq.u.soft.bellman.app.terminal}
 U^*(\bs) = \log \E_{\ba \sim p(\cdot|\bs)} \left [\exp \left ( r(\bs,\ba) + U^*(\ba(\bs)) \right )\right ]
\end{equation}
\begin{equation} \label{eq.q.soft.bellman.app.terminal}
 Q^*(\bs, \ba) = r(\bs, \ba) + \log \E_{\ba' \sim p(\cdot|\ba(\bs))} \left [\exp \left (  Q^*(\ba(\bs), \ba') \right) \right ]
\end{equation}
\end{subequations}
}

When the reference policy is uniform, it is easier to see how these are the `soft' version of the usual Bellman optimality equations: in this case we recognize the $\text{LogSumExp}$ function which can be interpreted as a soft version of the maximum function.

Finally, it is instructive to see the relation between the optimal action-value function $Q^*$ and the optimal policy $\pi^*$. Usually, the action is selected that has the maximum action-value, i.e. the argument of the maximum is used. In maximum entropy reinforcement learning, the optimal policy follows instead from the soft argument of the maximum, i.e.\
\begin{equation} \label{eq:policy.from.q}
 \pi^*(\ba|\bs) = \frac{\exp(Q^*(\bs,\ba))}{\E_{\ba' \sim p(\cdot|\bs)} \left [\exp \left (  Q^*(\bs, \ba') \right) \right ]}.
\end{equation}
%

\section{Proof of Reinforcement Learning Interpretation of Feynman--Kac Formula}
\label{sec:fkbellman}
In this section, we prove the assertions in Section \ref{sec:transform.fk}. For clarity, we begin by repeating the Feynman--Kac representation of the ground state \eqref{eq.fk.ground}
\begin{equation}
\varphi_0(\bs_t) = \E_\bbP
 \left[\exp\left(-\int_{t}^{t+T} \left (V({\bs_{t'}}) - E_0 \right ) dt' \right)\varphi_0({\bs_{t+T}})\right].
\end{equation}
For a small time step $T = \Delta t$, this implies
\begin{align}
\varphi_0(\bs) 
    &= \E_{\bbP} \left [ \exp(-(V(\bs)-E_0) \Delta t +\cO(\Delta t^2) ) \varphi_0 (\bs_{t+\Delta t}) \right] \\
    &= e^{r(\bs)+E_0 \Delta t} \E_{\bbP} \left [  \varphi_0 (\bs_{t+\Delta t})) \right] +\cO(\Delta t^2)  \label{eq.almost.todorov}
\end{align}
where the reward is defined as
\begin{equation}
 r(\bs) \equiv -V(\bs) \Delta t. 
\end{equation}

\subsection{Definition of the action space}
Although \eqref{eq.almost.todorov} already looks very similar to Todorov's linear equation for the desirability \eqref{eq.todorov.desirability}, we still have to rewrite the expectation over paths of length $\Delta t$ in terms of actions. That is, we replace the expectation over trajectories $\bbP$ with an expectation with respect to a `passive policy' $p_{\Delta t}(\cdot|\bs)$. This passive policy is a discrete time approximation of the path measure $\bbP$.

To define the policy, we first clarify what the available actions are. An action is either a move to another state $\bs'\neq\bs$ with $H_{\bs\bs'}\neq0$, or the trivial action $\ba_0$ of doing nothing, i.e. $\ba_0(\bs) = \bs$. Since $\bbP$ is characterized by the transition rates $\Gamma$ \eqref{eq.passive.dynamics}, the passive policy becomes
\begin{equation} \label{eq:passive.policy}
p_{\Delta t}(\ba|\bs) =
  \begin{cases}
   \Gamma_{\bs\to \ba(\bs)} \Delta t         & \ba \neq \ba_0 \\
   1 - \sum \Gamma_{\bs\to \ba(\bs)} \Delta t & \ba = \ba_0.
  \end{cases} 
\end{equation}
With this definition of the passive policy $p_{\Delta t}$, we can further rewrite \eqref{eq.almost.todorov} as
\begin{equation} \label{eq:really.todorov}
    \varphi_0(\bs) 
    = e^{r(\bs)+E_0 \Delta t} \E_{\ba\sim p_{\Delta t}(\cdot|\bs)} \left [ \varphi_0 (\ba(\bs)) \right] +\cO(\Delta t^2).
\end{equation}

\subsection{Soft Bellman optimality equation} \label{sec:discrete.fk}
We recognize this as the same equation as Todorov's linear equation for the desirability \eqref{eq.todorov.desirability}. Therefore, if we define the state-value function $U^*(\bs) = \log \varphi_0(\bs)$, we get the soft Bellman optimality equation \eqref{eq.u.soft.bellman}
\begin{equation} \label{eq:fk.u.bellman}
 U^*(\bs) = R^* + \log \E_{\ba \sim p_{\Delta t}(\cdot|\bs)} \left [\exp \left (r(\bs) + U^*(\ba(\bs)) \right )\right ]
\end{equation}
with $R^* = E_0 \Delta t$, which is illustrated in \Cref{fig:bellman}.  Therefore, using a small time step $\Delta t$, the Feynman--Kac formula is approximately equivalent to a reinforcement learning problem with discrete time steps $\Delta t$, an infinite time horizon and reward function $r(\bs)=-V(\bs)\Delta t$.



\subsection{The continuous time limit}
In continuous time, the actions taken by an agent cannot be specified by a policy. The reason is simple: the probability of taking an action in infinitesimal time is zero. Therefore, the control of an agent is specified with transition rates instead. \cite{todorov_efficient_2009} shows how to perform the continuous time limit in continuous state spaces. We extend this argument to discrete state spaces.

In continuous time, the control is specified by transition rates $\Gamma^{\theta}$ instead of a policy. For a small discrete time step $\Delta t$, the transition rates and policy are related by
\begin{equation}
\pi^{\theta}(\ba|\bs) =
  \begin{cases}
   \Gamma^{\theta}_{\bs\to \ba(\bs)} \Delta t{+\cO(\Delta t^2)}         & \text{if } \ba \neq \ba_0, \\
   1 - \sum \Gamma^{\theta}_{\bs\to \ba(\bs)} \Delta t{+\cO(\Delta t^2)}  & \text{if } \ba = \ba_0.
  \end{cases} 
\end{equation}
Comparing with \eqref{eq:passive.policy}, we see that the passive policy of course corresponds to $\Gamma^{\theta} = \Gamma$.

Next, we compute the relative entropy
\begin{equation}
 \mathcal{H}(\pi^{\theta}(\cdot | \bs)||p_{\Delta t}(\cdot | \bs)) = \Delta t\sum_{\bs'\neq \bs} \left(
    {\Gamma_{\bs\to \bs'}} 
    - \Gamma^{\theta}_{\bs\to \bs'}
    + \Gamma^{\theta}_{\bs\to \bs'} 
        \log\left(\frac{\Gamma^{\theta}_{\bs\to \bs'}}{\Gamma_{\bs\to \bs'}}\right) \right){+\cO(\Delta t^2)} .
\end{equation}
Now we can take the limit $\Delta t\to 0$. We arrive at a continuous time reinforcement learning problem, where the goal is to find the transition rates $\Gamma^{\theta}$ that maximize the objective \eqref{eq:holland.reward.discrete}.

{

\section{Training details} \label{sec:hparams}
\Cref{tab:hparams} lists the hyperparameters we used in all our experiments. Training follows the Soft Q-learning algorithm \citep{haarnoja_reinforcement_2017} with adaptions for discrete action spaces by \cite{christodoulou_soft_2019}. In contrast with \cite{christodoulou_soft_2019}, for simplicity we use only two neural networks: $Q = NN(\theta)$  and a target $\Bar{Q} = NN(\Bar{\theta}$). The policy is derived from $Q^*$ following \eqref{eq:policy.from.q}. For the continuous time formulation, we use the discrete time approximation described in \Cref{sec:discrete.fk} with a timestep $\Delta t = 10^{-4}$. This is because Soft Q-learning is not directly suitable for continuous time reinforcement learning.

To keep the architecture of the neural networks as simple as possible, we used a convolutional neural network with half-padding, so that the shape of the input is preserved. For a 6x6 Ising model, this means that the output of the neural network is also a 6x6 array of real numbers, representing the action-value of flipping the spin at that lattice site. For the continuous and non-terminal formulations, however, we also need the action-value of the trivial action $\ba_0$, i.e.\ not flipping any spin $\ba_0(\bs) = \bs$. From the relation $Q^*(\bs, \ba) = R^* + r(\bs) + \log \varphi_0(\ba(\bs))$ and the Schr\"odinger equation \eqref{eq:stochastic.schrodinger.2}, it follows that
$$
 \exp \left(Q^*(\bs, \ba_0) \right) = -\frac{1}{H_{\bs\bs} - E_0} \sum_{\ba\neq\ba_0} H_{\bs\bs'} \exp(Q^*(\bs, \ba))
$$
%
so that we do not need to represent $Q(\bs, \ba_0)$ separately. For this trick, we need an estimate of the ground state energy, and this is obtained from the energy validation of the wavefunction following from $Q$. This does not result in extra computation: the energy validation (done every $20$ episodes) needs to be done to keep track of training progress anyway.

Initial states are generated by randomly selecting half the spin sites and designating those spin up, the others spin down. Training starts after the replay buffer has been filled completely with experiences (states, actions, rewards, new states) following a uniform policy (i.e. every action has equal probability). For the terminal states formulation, we choose the entirely magnetized states to be terminal. Although these states are then terminal in the sense that $U^*=0$, regeneration of initial states is unnecessary. 

Every training episode, one batch of experiences is generated and saved to the replay buffer, after which one batch of experiences is sampled from the replay buffer $\mathcal{D}$. Then that batch is used to do one gradient step. The loss used in the gradient step derives from the soft Bellman optimality equation for $Q^*$. For the continuous time and non-terminal states formulation, the loss is the variance of the soft Bellman residue \citep{sutton_reinforcement_2018}
$$
 Var_{\bs, \ba\sim \mathcal{D}} \left[ Q(\bs, \ba) - r(\bs) - \log \E_{\ba' \sim p(\cdot|\ba(\bs))} \left [\exp \left (  \Bar{Q}(\ba(\bs), \ba') \right) \right ] \right ],
$$
i.e.\ the variance of the soft Bellman equation \eqref{eq.q.soft.bellman.app}. Note that the target network $\Bar{Q}$ is used in the exponential, and that the constant $R^*$ disappears. For the terminal states formulation, it is the mean squared error \citep{sutton_reinforcement_2018}
$$
 \E_{\bs, \ba \sim \mathcal{D}} \left[ \left ( Q(\bs, \ba) - r(\bs) - \log \E_{\ba' \sim p(\cdot|\ba(\bs))} \left [\exp \left (  \Bar{Q}(\ba(\bs), \ba') \right) \right ] \right )^2\right ],
$$
i.e.\ the mean squared error of the soft Bellman equation \eqref{eq.q.soft.bellman.app.terminal}. The subscripts ${\bs, \ba \sim \mathcal{D}}$ indicate that these losses are computed using batches sampled from the replay buffer $\mathcal{D}$. In contrast, the expectations inside the logarithm can be computed exactly, since one neural network call outputs $\Bar{Q}(\ba(\bs), \ba')$ for all $\ba'$ and the transition probabilities $p(\cdot|\ba(\bs))$ are known.

Training is done for 2000/4500/4500 episodes for the continuous/non-terminal/terminal formulations respectively, taking around 20-40 minutes on a 12 GB GPU.

\begin{table}[]
    
    \centering
    \begin{tabular}{c|c}
    Parameter & Value  \\
    \hline
    optimizer & Adam \citep{kingma_adam_2017} \\
    learning rate & $10^{-3}$ \\
    learning rate decay & $\times0.99$ every 10 episodes\\
    batch size & 4096 \\
    replay buffer size & 65536 \\
    target update interval & 20 \\
    hidden layers  & 3 \\
    convolutional channels & 64 \\
    convolutional kernel size & 3 \\
    convolutional stride & 1 \\
    convolutional padding & circular, 1 \\
    nonlinearity & ReLU \\
    \end{tabular}
    \caption{Hyperparameters}
    \label{tab:hparams}
\end{table}

}
\end{document}